\documentclass[dvipsnames]{article}
\usepackage[pass]{geometry}

\usepackage{xcolor,attrib}
\usepackage[sorting=none,backend=biber]{biblatex}
\usepackage{hyperref}
\usepackage{xurl}
\hypersetup{breaklinks=true,colorlinks=true,urlcolor=Blue,citecolor=Blue}
\usepackage{authblk,afterpage}
\usepackage{palatino,multicol,orcidlink}

\title{Guarding against artificial intelligence--hallucinated citations}
\usepackage{etoolbox}
\makeatletter
\providecommand{\subtitle}[1]{% add subtitle to \maketitle
  \apptocmd{\@title}{\par\vskip.5\baselineskip {\Large #1 \par}}{}{}
}
\makeatother
\subtitle{The case for full-text reference deposit}
\author{Alex Glynn, MA\,\orcidlink{0000-0002-3027-7276}}
\affil{Kornhauser Health Sciences Library\\University of Louisville\\Louisville, KY\\United States of America}
\affil{\href{mailto:alex.glynn@louisville.edu}{alex.glynn@louisville.edu}}
\affil{\href{https://www.academ-ai.info/}{academ-ai.info}}

\date{Submitted to arXiv March 24, 2025}

\begin{document}
\maketitle
\begin{abstract}
\noindent\sloppy The tendency of generative artificial intelligence (AI) systems to
``hallucinate'' false information is well-known; AI-generated citations
to non-existent sources have made their way into the reference lists of
peer-reviewed publications. Here, I propose a solution to this problem,
taking inspiration from the Transparency and Openness Promotion (TOP)
data sharing guidelines, the clash of generative AI with the American
judiciary, and the precedent set by submissions of prior art to the
United States Patent and Trademark Office. Journals should require
authors to submit the full text of each cited source along with their
manuscripts, thereby preventing authors from citing any material whose
full text they cannot produce. This solution requires limited additional
work on the part of authors or editors while effectively immunizing
journals against hallucinated references.
\end{abstract}

\newgeometry{margin=.75in, columnsep=.2in}

\begin{multicols}{2}
\noindent On March 4, 2024, \emph{PLoS One} published an article by Shoukat et al.\
comparing blended learning with traditional instruction for
undergraduates in Faisalabad, Pakistan.\supercite{shoukat2024} Within the
same month, commenters on PubPeer raised concerns regarding the
article's reference list.\supercite{pubpeer} The first clear irregularity
was the extraneous phrase ``Regenerate response,'' which appeared in
reference \#62. This phrase was formerly the label of a button on the
ChatGPT interface. In at least 50 instances, authors have apparently
copied the label along with the chatbot's output and pasted it into
their subsequently published manuscripts.\supercite{acai} On further
investigation into the \emph{PLoS One} article, PubPeer commentors were
unable to find sources matching several of the references, including
\#62, concluding that they likely did not exist.\supercite{pubpeer}
Generative artificial intelligence (AI) tools based on large language
models (LLMs) have a known tendency to ``hallucinate'' false
information\supercite{nist2024}, particularly bibliographic
references.\supercite{mugaanyi2024}

On April 18, 45 days after publication, the article was retracted. The
editors cited the phrase ``Regenerate response'' as evidence of
``potential undisclosed use of an AI tool'' and reported that 18 of the
article's 76 references could not be identified.\supercite{shoukat2024e}
Had their authenticity been checked during peer review or other
editorial processes, the study may never have been published. However,
editorial verification is an inefficient solution to this problem.
Proving definitively that a hallucinated reference does not exist is,
like many negative assertions, time-consuming at best and impossible at
worst. Proving that a real reference does exist is typically a trivial
endeavor.

An instructive parallel comes from the legal world. In 2023, two
attorneys were sanctioned in the Southern District of New York for
``submitt{[}ing{]} non-existent judicial opinions with fake quotes and
citations created by the AI tool ChatGPT''; \emph{Mata v.~Avianca,
Inc.}, 678 F.~Supp.~3d 443, 448 (S.D.N.Y. 2023). The non-existent
opinions were cited in the Plaintiff's Affirmation in Opposition. When
first the Defendent's counsel and then the Court itself were unable to
locate the cited decisions, the Plaintiff's counsel were instructed to
provide copies of them in full text. The attorneys were able to provide
only ``excerpts,'' also hallucinated by ChatGPT, which themselves
contained further citations to non-existent cases, among various other
legal and logical flaws.

Two primary factors contributed to the discovery of the hallucination in
this case. Firstly, the adversarial legal system, unlike academic peer
review, strongly incentivizes lawyers to find and read the full text of
cases cited by the opposing counsel. Secondly, when neither the Defense,
nor the Court was able to do so, the Plaintiff's counsel were required
to provide the full text themselves. Had the cases actually existed,
this would have been an extremely simple task.

Among the ``{[}m{]}any harms'' caused by the submisison of fake
opinions, the Court cited the following:

\begin{quote}
The opposing party wastes time and money in exposing the deception. The
Court's time is taken from other important endeavors.

\attrib{\emph{Id.}}
\end{quote}
Searching fruitlessly for non-existent journal articles is no better use
of time for editors and peer reviewers than searching fruitlessly for
non-existent case law is for courts and law firms. In academic
publishing, rather than attempting to require or incentivize peer
reviewers to verify bibliographic references, editors could require
authors, to the extent permissible by applicable copyright law, to
submit the full text of each reference as part of the submission
process.

Further precedent comes from the increasingly common practice of data
deposit. The Transparency and Openness Promotion (TOP) guidelines
provide a set of data deposit standards for
journals.\supercite{nosek2015} To achieve level II or III data
transparency, journals must require that original research data
associated with submitted articles be desposited in a trusted
repository, such as Dataverse or Dryad, in the interest of
reproducibility. As of March 12, 2025, 320 journals have achieved this
standard.\supercite{stodden2016}

Depositing full-text references would not be as resource-intensive as
depositing research data. The full texts of cited articles need not be
\emph{publicly} available; they would only need to be shown to the
editors and peer reviewers to verify their authenticity and, as an added
benefit, be available for the reviewers' use while appraising the
submitted article.

It should not put undue strain on authors to submit copies of their
sources in full text. Since responsible researchers do not cite articles
based on their abstracts alone, the citing author should already have
access to the full text. For longer works, such as books and monographs,
only the relevant chapters or sections need be provided. Items available
only in physical formats could be scanned or photographed. Thanks to the
successful implementation of Inter-Library Loan services and other
resource sharing initiatives, accessing and sharing full-text sources is
largely a solved logistical problem.

One foreseeable impediment to implementing my recommendation is that
sending the full text of closed access
sources to journal editors may infringe on the copyright of those
sources. In the United States, exceptions to copyright exist on the
grounds of fair use (17 U.S.C. § 107), which may protect the submission
of full-text references to journals; American courts have found in favor
of fair use defenses in similar scenarios. For example, a 2013 ruling
upheld the fair use defense for a law firm making copies of scientific
articles (non-patent literature, or NPL) for use as evidence in patent
filings; \emph{Am. Inst. of Physics v. Winstead PC}, Civil Action
No.~3:12-CV-1230-M (N.D. Tex. 2013). The Court found that ``the
Defendants' copying of NPL serves an evidentiary function'' and
``facilitates an efficient patent process'' and that ``it is not merely
the content of the NPL article that matters, but its existence in the
available literature.'' Likewise, the submission of full-text references
to journals would serve an evidentiary function---verifying the
references' existence in the available literature---and facilitate an
efficient peer review process.

However, an international publishing industry has to function within
many different legal
systems, including those lacking the exceptions to copyright protections found in American law;
in at least some cases, exceptions to the full text deposit requirement will likely be necessary.
Following TOP data transparency level II, journals could require that
exceptions to the full-text deposit requirement be identified and
justified on submission.\supercite{nosek2015} Only these exceptions would
then need to be investigated by editors or peer reviewers. Even if the full text cannot be provided, journals could
require that all citations include persistent identifiers or links to
records in bibliographic databases or library catalogues. Such links
still require validation since they can be generated easily by LLMs, but
testing a link is much less time-consuming than validating a plain
bibliographic reference.

It has always been possible for unscrupulous researchers to cite
non-existent references, gambling on the assumption that editors cannot
or will not verify their authenticity. However, the emergence of popular
and affordable generative AI tools has enabled researchers to cite
non-existent references not only with unprecedented ease, but even
without intent. So long as AI tools capable of hallucination are in use,
this problem is likely to continue. The solution proposed here would
prevent authors from citing non-existent sources; to continue doing so, authors would need to fabricate entire articles or create
counterfeit bibliographic databases, measures which, while certainly not
impossible, lack the expediency that attracts unscrupulous authors to
generative AI and certainly rule out any accidental fabrication.

\section{Declarations}\label{declarations}

\subsection{Conflict of Interest}\label{conflict-of-interest}

The author is the founder and maintainer of Academ-AI (reference 3), a
non-revenue generating project. The author declares no other conflict of
interest, financial or otherwise.

\subsection{Funding}\label{funding}

The author received no specific funding for this work.

\printbibliography

@article{shoukat2024e,
  author    = {{{The PLoS One Editors}}},
  journal   = {PLoS One},
  title     = {Retraction: A comparative analysis of blended learning and traditional instruction: Effects on academic motivation and learning outcomes},
  year      = {2024},
  issn      = {1932-6203},
  month     = apr,
  number    = {4},
  pages     = {e0302484},
  volume    = {19},
  doi       = {10.1371/journal.pone.0302484},
  publisher = {Public Library of Science (PLoS)},
}

@article{shoukat2024,
  author  = {Shoukat, Rizwan and Ismayil, Iskander and Huang, Qibing and Oubibi, Mohamed and Younas, Muhammad and Munir, Rizwan},
  journal = {PLoS One},
  title   = {A comparative analysis of blended learning and traditional instruction: Effects on academic motivation and learning outcomes},
  year    = {2024},
  issn    = {1932-6203},
  month   = mar,
  number  = {3},
  pages   = {e0298220},
  volume  = {19},
  doi     = {10.1371/journal.pone.0298220},
}

@techreport{nist2024,
  author      = {Raimondo, Gina M. and Locascio, Laurie E.},
  institution = {{National Institute of Standards and Technology}},
  title       = {Artificial intelligence risk management framework generative artificial intelligence profile},
  date        = {2024-07},
  doi         = {10.6028/nist.ai.600-1},
  publisher   = {U.S. Department of Commerce},
}

@article{nosek2015,
  author    = {Nosek, B. A. and Alter, G. and Banks, G. C. and Borsboom, D. and Bowman, S. D. and Breckler, S. J. and Buck, S. and Chambers, C. D. and Chin, G. and Christensen, G. and Contestabile, M. and Dafoe, A. and Eich, E. and Freese, J. and Glennerster, R. and Goroff, D. and Green, D. P. and Hesse, B. and Humphreys, M. and Ishiyama, J. and Karlan, D. and Kraut, A. and Lupia, A. and Mabry, P. and Madon, T. and Malhotra, N. and Mayo-Wilson, E. and McNutt, M. and Miguel, E. and Paluck, E. Levy and Simonsohn, U. and Soderberg, C. and Spellman, B. A. and Turitto, J. and VandenBos, G. and Vazire, S. and Wagenmakers, E. J. and Wilson, R. and Yarkoni, T.},
  journal   = {Science},
  title     = {Promoting an open research culture},
  year      = {2015},
  issn      = {1095-9203},
  month     = jun,
  number    = {6242},
  pages     = {1422--1425},
  volume    = {348},
  doi       = {10.1126/science.aab2374},
  publisher = {American Association for the Advancement of Science (AAAS)},
}

@online{stodden2016,
  addendum     = {Modified March 12, 2025},
  author       = {Stodden, Victoria and Boycan, Ella and Mellor, David Thomas and Lowrey, Olivia and Esposito, Jolene and DeHaven, Alexander C.},
  date         = {2016-06-20},
  doi          = {10.17605/OSF.IO/KGNVA},
  organization = {Open Science Framework},
  title        = {TOP Resources - Evidence and Practices},
}

@online{pubpeer,
  commentator  = {Cabanac, Guillaume and Abalkina, Anna},
  organization = {PubPeer},
  title        = {A comparative analysis of blended learning and traditional instruction: Effects on academic motivation and learning outcomes},
  url          = {https://www.pubpeer.com/publications/7F01171F32B0421DF37EE8D1B49C04},
}

@online{acai,
  organization = {Academ-AI},
  title        = {Academ-AI},
  url          = {https://www.academ-ai.info/},
  urldate      = {2025-03-14},
}

@article{mugaanyi2024,
  author   = {Mugaanyi, Joseph and Cai, Liuying and Cheng, Sumei and Lu, Caide and Huang, Jing},
  journal  = {J Med Internet Res},
  title    = {Evaluation of {Large} {Language} {Model} {Performance} and {Reliability} for {Citations} and {References} in {Scholarly} {Writing}: {Cross}-{Disciplinary} {Study}},
  year     = {2024},
  month    = apr,
  number   = {1},
  pages    = {e52935},
  volume   = {26},
  doi      = {10.2196/52935},
}

\end{multicols}

\end{document}